\documentclass[conference]{IEEEtran}
\ifCLASSINFOpdf
\else
\fi

\usepackage{psfrag,epsfig}
\usepackage{amsmath,amssymb}
\usepackage{algorithm,algorithmic}
\usepackage{subfig}

\usepackage[utf8]{inputenc}

\newcommand{\be}{\begin{equation}}
\newcommand{\ee}{\end{equation}}
\newcommand{\beqy}{\begin{eqnarray}}
\newcommand{\eeqy}{\end{eqnarray}}
\newcommand{\beqynn}{\begin{eqnarray*}}
\newcommand{\eeqynn}{\end{eqnarray*}}

\newcommand{\ba}{\begin{array}}
\newcommand{\ea}{\end{array}}
\newcommand{\bmx}{\begin{bmatrix}}
\newcommand{\emx}{\end{bmatrix}}
\newcommand{\bsmx}{\left[\begin{smallmatrix}}
\newcommand{\esmx}{\end{smallmatrix}\right]}
\newcommand{\bmxc}[1]{\left[\begin{array}{@{}#1@{}}}
\newcommand{\emxc}{\end{array}\right]}

\newcommand{\bt}[1]{\begin{tabular}{#1}}
\newcommand{\et}{\end{tabular}}

\newcommand{\bc}{\begin{center}}
\newcommand{\ec}{\end{center}}
\newcommand{\ben}{\begin{enumerate}}
\newcommand{\een}{\end{enumerate}}
\newcommand{\bi}{\begin{itemize}}
\newcommand{\ei}{\end{itemize}}

\newcommand{\X}{{\boldsymbol{X}}}

\newcommand{\x}{{\boldsymbol{x}}}

\begin{document}
%
\title{Distributed Cooperative Localization in Wireless Sensor Networks without NLOS Identification}

\author{\IEEEauthorblockN{Siamak Yousefi\IEEEauthorrefmark{1},
Xiao-Wen Chang\IEEEauthorrefmark{2}, and
Benoit Champagne\IEEEauthorrefmark{1} }
\IEEEauthorblockA{\IEEEauthorrefmark{1}Department of Electrical and Computer Engineering,
McGill University, Montreal, Quebec, H3A 0E9, Canada \\
} 
\IEEEauthorblockA{\IEEEauthorrefmark{2}School of Computer Science,
McGill University, Montreal, Quebec, H3A 0E9, Canada\\
Email: siamak.yousefi@mail.mcgill.ca; chang@cs.mcgill.ca; benoit.champagne@mcgill.ca}
 }


%


\maketitle

\begin{abstract}
In this paper, a 2-stage robust distributed algorithm is proposed for cooperative sensor network localization using time of arrival (TOA) data without identification of non-line of sight (NLOS) links.
In the first stage, to overcome the effect of outliers, a convex relaxation of the Huber loss function is applied so that by using iterative optimization techniques, good estimates of the true sensor locations can be obtained.
In the second stage, the original (non-relaxed) Huber cost function is further optimized to obtain refined location estimates based on those obtained in the first stage.
In both stages, a simple gradient descent technique is used to carry out the optimization.
Through simulations and real data analysis, it is shown that the proposed convex relaxation generally achieves a lower root mean squared error (RMSE) compared to other convex relaxation techniques in the literature.
Also by doing the second stage, the position estimates are improved and we can achieve an RMSE close to that of the other distributed algorithms which know \textit{a priori} which links are in NLOS.
\end{abstract}

%
\IEEEpeerreviewmaketitle

\section{Introduction}
Wireless sensor network (WSN) localization has received great attention in recent years due to the large number of applications requiring accurate location information \cite{Sayed}.
Since the global positioning system (GPS) is not a reliable technology for localization of sensors in indoor place or dense urban areas,
range measurements between pairs of neighbouring nodes, including sensors and anchors, may be used for the purpose of localization.
Among the different technologies, ultra-wide band (UWB) signalling can yield accurate time of arrival (TOA) measurements in line of sight (LOS) scenarios, from which the range information can be extracted.

The localization based on these measurements can be carried out for the entire network in a centralized or a distributed fashion.
Among the popular centralized algorithms are semi-definite programming (SDP) \cite{Biswas_SDP} and second-order cone programming (SOCP) \cite{Tseng_SOCP} convex relaxations.
Distributed algorithms have also been proposed, including distributed SOCP \cite{Ghasem_SOCP}, the iterative parallel projection method (IPPM) \cite{Jia_IPPM} and other localization approaches that alternate between convex and non convex optimization problems \cite{Abramo} \cite{Distributed_Localization_Zhu}.

However, these approaches only consider the case where the pairwise range measurements are made under LOS condition.
In practice, LOS measurements are limited and many links will face a non-line of sight (NLOS) condition.
Due to the NLOS, the TOA measurements become positively biased \cite{Guvenc}, and consequently, 
the aforementioned techniques perform unsatisfactorily if the NLOS effects are not mitigated properly.

In many of the localization techniques, the NLOS links have to be identified first.
Various methods have been proposed for the identification of NLOS links in non-cooperative networks (see in \cite{Guvenc} and the references therein), among which several are especially tailored for UWB applications  \cite{Identify-UWB-Marano}, \cite{Identification_Buehrer}.
After detecting the NLOS links through a suitable technique, the effect of NLOS error can be mitigated using different optimization techniques.
A summary of the non-cooperative TOA-based NLOS mitigation techniques is given in \cite{Guvenc}.
For cooperative localization, extension of the centralized SDP relaxation and the distributed IPPM to NLOS scenarios are considered in \cite{Vaghefi} and \cite{Jia_IPPM_NLOS}, respectively.

NLOS identification remains however challenging for a large WSN with several pairwise measurements.
Therefore, in many applications, it is impractical to assume that all the NLOS links can be identified accurately.
In \cite{Vaghefi_Unidentified}, an SDP relaxation is considered for non-cooperative localization in NLOS without prior detection of NLOS links. 
Although this technique is robust against NLOS errors, the computations need to be done centrally, thus it can not scale with the size of the network, and its extension to a distributed implementation remains an open topic for further study.
In \cite{Gholami_POCS}, a distributed cooperative projection onto convex sets (POCS) is employed to estimate the location of sensors, which is shown to be robust against NLOS errors.
However, if only a portion of the measurements are affected by NLOS errors, the performance of POCS is far from being optimal.
Another approach for robust estimation against outliers without prior outlier detection is to use Huber loss function, which offers a trade-off between $l_1$ and $l_2$ norm minimizations \cite{Huber_Robust}.
In contrast to POCS, localization based on Huber cost function can achieve a good result only if it is well initialized and if a moderate or small portion of the measurements are contaminated by large errors, otherwise it may not necessarily give a good estimate.

In this paper, to obtain accurate sensor location estimates under various NLOS scenarios, we propose a 2-stage algorithm based on Huber M-estimation for distributed cooperative localization in the presence of unidentified NLOS links.
In the first stage, a convex relaxation similar to that in \cite{Abramo} is applied on the Huber cost function and sensor locations are then estimated iteratively. 
Since the performance may not be close to optimal when the ratios of NLOS to LOS links is low, in the second stage, the original Huber cost function is minimized iteratively with a suitable choice of tuning parameter.
For the iterative optimization in both stages, we use a simple gradient descent technique since it can be easily implemented in a distributed manner.
Through simulations, we first show that the proposed convex relaxation gives a robust estimate in different NLOS scenarios.
Furthermore, we show that the position estimates are generally improved in the second stage as we minimize the original Huber cost function.
The robustness of our algorithm to outliers is also evaluated by using a real set of sensor measurements obtained by the measurement campaign in \cite{Patwari_Camp}.


\section{System Model and Problem Formulation} \label{Sec:Problem}
\subsection{System Model}
We consider a sensor network consisting of $N$ sensor nodes with unknown locations denoted by $\x_i \in \mathbb{R}^2$, for $i=1,\ldots,N$, and $M$ anchors with known locations $\x_i \in \mathbb{R}^2$, for $i = N+1, \ldots, N+M$.
We define ${\cal S}$ as the set of all index pairs $(i,j)$ of all the neighbouring nodes that can communicate with each other, where we let $i<j$ to avoid repetition.
We also define ${\cal S}_i$ as the index set of all the neighbouring nodes of the $i$-th sensor.
We assume that a range measurement is obtained between each pair of neighbouring nodes with $(i,j) \in {\cal S}$.
For accurate TOA-based ranging, we either assume that the nodes are precisely synchronized over the network, or that the two-way ranging (TWR) protocol is employed to remove the clock error in the TOA measurements \cite{Dardari}.
The complete set of range measurements are modelled by the following equations:
\be
\label{Range}
r_{ij}=
\begin{cases}
d_{ij} + n_{ij}, \quad  &(i,j) \in {\cal L} \\
d_{ij} + b_{ij} + n_{ij}, \quad  &(i,j) \in {\cal N} 
\end{cases}
\ee
where we define $d_{ij}= \|\x_i-\x_j \|$, along with the sets 
\begin{align}
{\cal L} &= \{ (i,j) \in {\cal S}: \text{LOS link between $i$-th and $j$-th node} \} \nonumber \\
{\cal N} &= \{ (i,j) \in {\cal S}: \text{NLOS link between $i$-th and $j$-th node} \} \nonumber
\end{align}
so that ${\cal S}= {\cal L} \cup {\cal N} $.
The measurement noise terms $n_{ij}$ are independent and identically distributed random variables with zero-mean and known variance $\sigma_n^2$.
The terms $b_{ij}$ are the NLOS biases between the corresponding pair of nodes indexed by $(i,j) \in {\mathcal N}$.
In the literature, the NLOS biases have been modelled differently depending on the environment and wireless channel, for instance, exponential \cite{Identify-UWB-Marano} or uniform \cite{Identification_Buehrer} distributions are generally used.
In this work, however, we do not assume any specific knowledge about the statistics of $b_{ij}$, such as its mean and variance.
Furthermore, no \emph{a priori} knowledge about the status of a link, i.e., whether it is NLOS or LOS, is assumed to be available.


\subsection{Problem Formulation}
\begin{figure*}[htbp]
\centering
\subfloat[] {
\includegraphics[width=55mm,height=55mm]{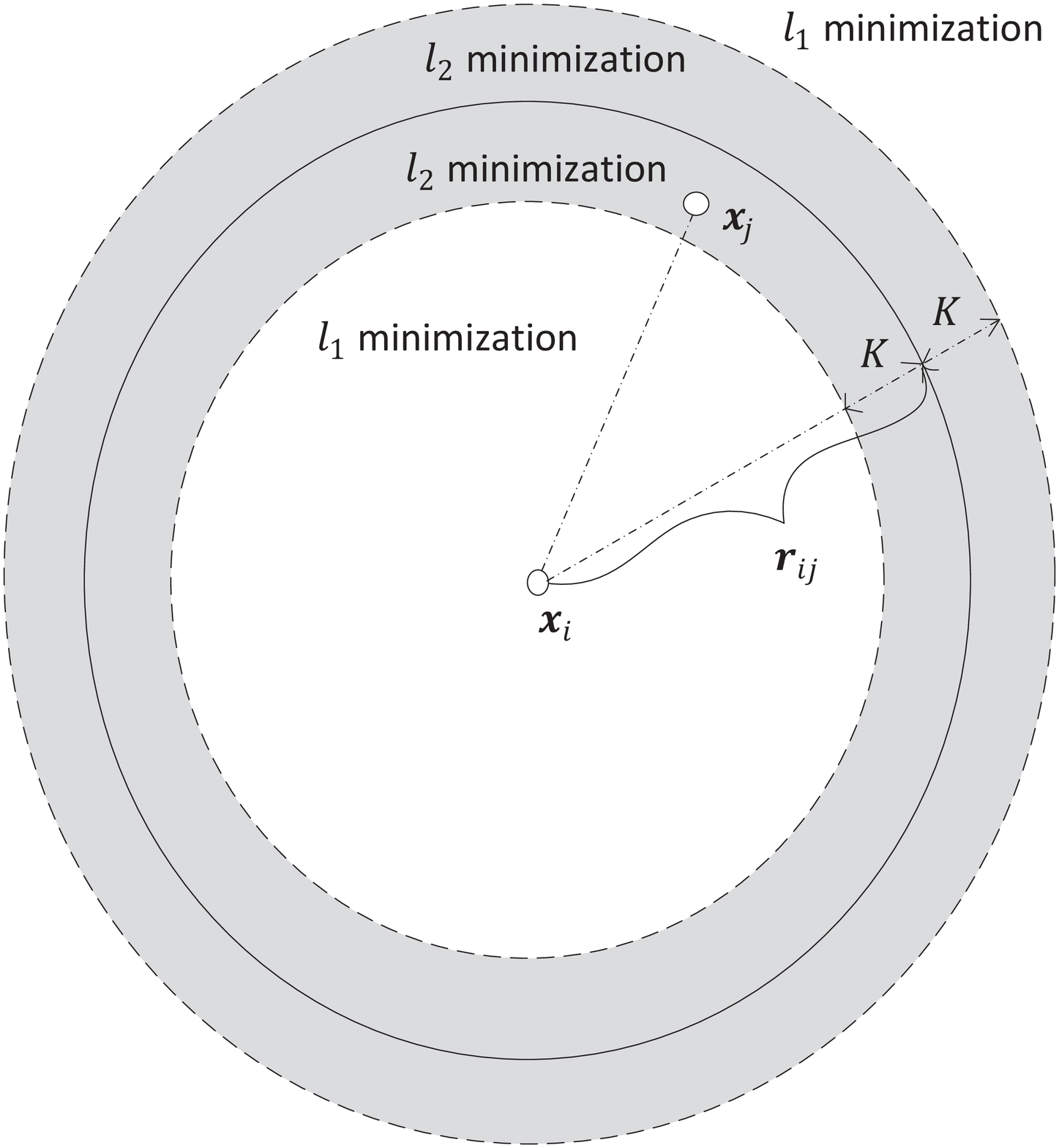}
\label{fig.:node_msd_trans_6nodes}
}\quad \quad
\subfloat[] {
\includegraphics[width=55mm,height=55mm]{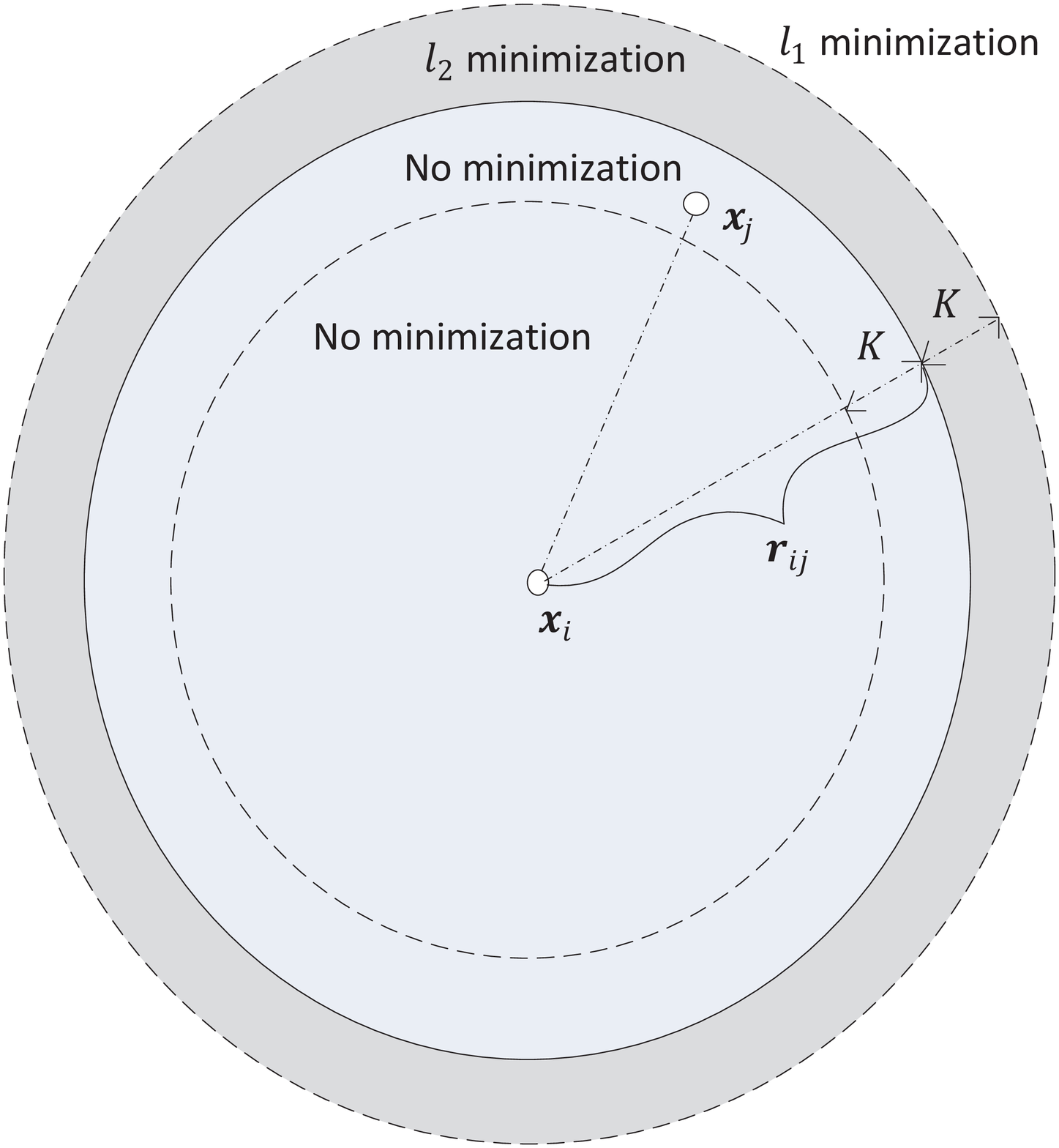}
\label{fig.:node_emse_trans_6nodes}
}
\caption{Illustration of two nodes and their pairwise range measurement. The regions where $l_1$ and $l_2$ norm minimization are implemented: (a) Original Huber cost function; (b) Proposed convex Huber cost function.}
\label{fig:Huber_Region}
\end{figure*}

The aim is to find estimates of the $N$ unknown sensor positions $\x_i$, denoted as $\hat{\x}_i$, such that the corresponding errors in the estimated locations are small, ideally unbiased with small variances. 
Throughout this work we denote $\X=[\x_1,\x_2,\ldots,\x_N] \in \mathbb{R}^{2\times N}$ as the unknown to be estimated.

If there are no NLOS biases, due to the zero-mean Gaussian noise assumption, it can be easily shown that the maximum likelihood estimation (MLE) is equivalent to the $l_2$ norm minimization, so the cost function to be minimized is
\be
\label{Cost_NLS}
f(\X) =   \sum_{(i,j)\in {\cal S}  }^{}  \Big( \|\x_j-\x_i \|-r_{ij}  \Big) ^2 
\ee
which is a non-convex nonlinear least squares (NLS) problem with respect to $\X$ \cite{Biswas_SDP}.
Since the NLOS biases exist in some measurements but cannot be identified, using $l_2$ norm minimization might not yield robust estimates. 
In the presence of outliers, Huber cost function provides a suitable replacement for $l_2$ norm minimization, by interpolating between $l_2$ and $l_1$ norm minimizations.
Therefore, instead of \eqref{Cost_NLS} it is preferred to minimize
\be \label{Huber_Cost}
g( {\X} ) = \sum_{(i,j) \in {\cal S} } \rho \Big(  \| \x_i  - \x_j \| - r_{ij}  \Big)
\ee
where $\rho(\cdot)$ is the continuous and differentiable Huber function:
\be
\label{Huber}
\rho(u_{ij})=
\begin{cases}
u_{ij}^2 , \quad &| u_{ij} | < K \\
2K| u_{ij} | - K^2 , \quad &|u_{ij}| \geq K \\
\end{cases}
\ee
The argument $u_{ij}= \|\x_i - \x_j \| - r_{ij}$, and $K$ is a fixed parameter which is chosen to be proportional to $\sigma_n$, e.g., $K=\alpha \sigma_n$ and $1.5 \leq \alpha \leq 2$ \cite{Huber_Robust}.

%

Although the Huber cost function is convex with respect to its argument, due to the non-convex nature of the range measurements with respect to the position coordinates, the final function is non-convex and hence initialization is crucial.
Furthermore, the Huber M-estimation can perform well only if a small or moderate portion of the measurements are affected by outliers, otherwise it may not achieve a good estimation result.
Therefore, in the following we propose a 2-stage algorithm that is robust in any NLOS scenario.

\section{Robust Distributed Algorithm} \label{Sec:Algorithm}
In this section, we first propose a convex relaxation of the Huber cost function.
After converging to some stationary points, we then try to minimize the Huber cost function.

\subsection{Stage I: Convex Relaxation}
A convex relaxation of the nonlinear least square problem in \eqref{Cost_NLS} has been proposed in \cite{Abramo} in the form of
\be
\label{Cost_Convex}
\tilde{f}(\X)  =  \sum_{(i,j)\in {\cal S} }^{}  \Big( (\|\x_j-\x_i \|-r_{ij})_{+} \Big) ^2  \\
\ee
where
\begin{equation*}
  (\|\x_j-\x_i \|-r_{ij})_{+} = \begin{cases}
               0  ,             & \|\x_j-\x_i \| \leq r_{ij}    \\
               \|\x_j-\x_i \|-r_{ij},  & \|\x_j-\x_i \| > r_{ij}  \\
           \end{cases}
\end{equation*}
Further explanations about the convexity of this cost function are given in \cite{Distributed_Localization_Zhu}.
The concept of this relaxation is similar to POCS proposed first in \cite{POCS} and considered for cooperative localization in \cite{Gholami_POCS}.
Here, however, we propose to minimize
\be \label{Huber_Cost_Convex}
\tilde{g}( {\X} ) = \sum_{(i,j) \in {\cal S} } \tilde{\rho} \Big( \| \x_i  - \x_j \| - r_{ij}  \Big)
\ee
where $\tilde{\rho}(\cdot)$ is the convex relaxation of the Huber function with respect to $\X$, which is defined as
\be
\label{Huber_Convex}
\tilde{\rho}  (u_{ij})=
\begin{cases}
0, \quad & \|\x_i - \x_j \| \leq r_{ij} \\
u_{ij}^2,  \quad &  r_{ij} \!<\! \|\x_i - \x_j \| \!<\! r_{ij}  + K_1 \\
2K_1 u_{ij}  - K_1^2, \quad & \|\x_i - \x_j \|  \geq r_{ij} + K_1 \\
\end{cases}
\ee
where $K_1=\alpha_1 \sigma_n$ is the parameter of the Huber loss function.
The geometric interpretation of the original and relaxed Huber cost functions are illustrated in Fig. \ref{fig:Huber_Region} for the area between two nodes.
Simulation result shows that, in many cases, this convex relaxation is more robust against large negative errors and gives a lower MSE for the network compared to the one in \eqref{Cost_Convex} or the cooperative POCS \cite{Gholami_POCS}.

The iterative gradient descent method for updating the position estimates can be stated at each node as
\be \label{Iterative_Huber_Relaxed}
\x_i^{(l+1)} = \x_i^{(l)} - \mu_1\sum_{j \in {\cal S}_i} \frac{\partial \tilde{\rho}(u_{ij})} {\partial \x_i^{(l)}} , \quad i = 1,\ldots, N
\ee
where $\mu_1$ is a suitable step size and for every $j \in {\cal S}_i$
\be
\frac{\partial \tilde{\rho}(u_{ij})} {\partial \x_i^{(l)}} =
\begin{cases}
0 ,  &\|\x_i^{(l)} - \x_j^{(l)} \| \leq r_{ij} \nonumber  \\
2u_{ij}^{(l)} \frac{\x_i^{(l)}-\x_j^{(l)}}{\|\x_i^{(l)} - \x_j^{(l)} \|}  , &r_{ij} \!<\! \|\x_i^{(l)} - \x_j^{(l)} \| \!<\! r_{ij}  + K_1 \nonumber \\
2K_1 \frac{\x_i^{(l)}-\x_j^{(l)}}{\|\x_i^{(l)} - \x_j^{(l)} \|} , &\|\x_i^{(l)} - \x_j^{(l)} \|  \geq r_{ij} + K_1  \nonumber \\
\end{cases}
\ee
After updating its location estimate using \eqref{Iterative_Huber_Relaxed}, each sensor sends the result to its neighbours.
Therefore, every sensor uses the current estimate about its own position, the known positions of its neighbouring anchors, and the updated positions of its neighbouring sensors to find a new estimate of its position. 
After convergence, we find an estimate of $\X$ which is the global minimum of the cost function in \eqref{Huber_Cost_Convex}.
The stopping criteria are either the maximum number of iterations or when the estimates of sensor positions at two consecutive iterations are smaller than a threshold, i.e., $\|\x_i^{(l+1)} - \x_i^{(l)} \| \leq \nu_1$ for all $i=1,\ldots,N$. 
The position estimates obtained at this stage are close to optimal if most of the measurements are NLOS.
However, in other scenarios, these estimates may not be close to optimal, and minimizing \eqref{Huber_Cost} will give a better estimate as will be explained in the sequel.

\subsection{Stage II: Position Refinement} 
At this stage, we try to minimize the original Huber cost function in \eqref{Huber_Cost}.
The iterative gradient decent steps at each sensor node $\x_i$ is 
\be
\x_i^{(l+1)} = \x_i^{(l)} - \mu_2\sum_{j\in {\cal S}_i} \frac{\partial \rho(u_{ij}) } {\partial \x_i^{(l)}} , \quad i = 1,\ldots, N
\ee
where $\mu_2$ is a suitable step size and for every $j \in {\cal S}_i$
\be
\frac{\partial  \rho(u_{ij}) } {\partial \x_i^{(l)}} = 
\begin{cases}
2u_{ij}^{(l)} \frac{\x_i^{(l)}-\x_j^{(l)}}{\|\x_i^{(l)} - \x_j^{(l)} \|}  ,\quad \Big| \|\x_i^{(l)} - \x_j^{(l)} \| - r_{ij} \Big| < K_2\\
2K_2 \frac{ \x_i^{(l)} - \x_j^{(l)} } {\| \x_i^{(l)} - \x_j^{(l)} \| } , \quad \Big| \|\x_i^{(l)} - \x_j^{(l)} \| - r_{ij} \Big| \geq K_2
\end{cases} \nonumber
\ee
and $K_2=\alpha_2 \sigma_n$ is the parameter of the Huber cost function.
The algorithm continues iteratively for a limited number of iterations similar to the first stage until convergence, i.e., $\|\x_i^{(l+1)}- \x_i^{(l)} \| \leq \nu_2$, for all $i=1,\ldots, N$.

\begin{figure*}[htbp]
\centering
\subfloat[]{
\includegraphics[width=51mm,height=39mm]{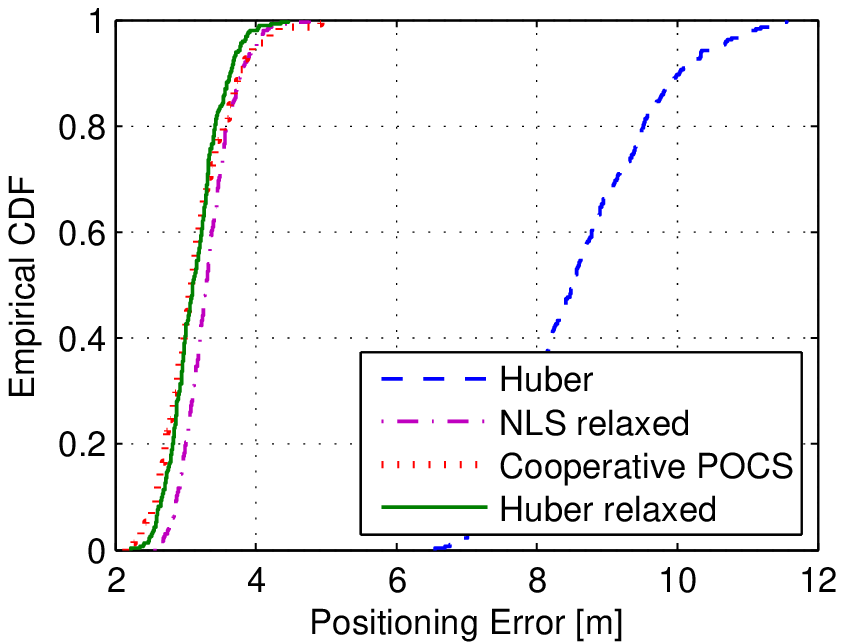}
}~
\subfloat[]{
\includegraphics[width=51mm,height=39mm]{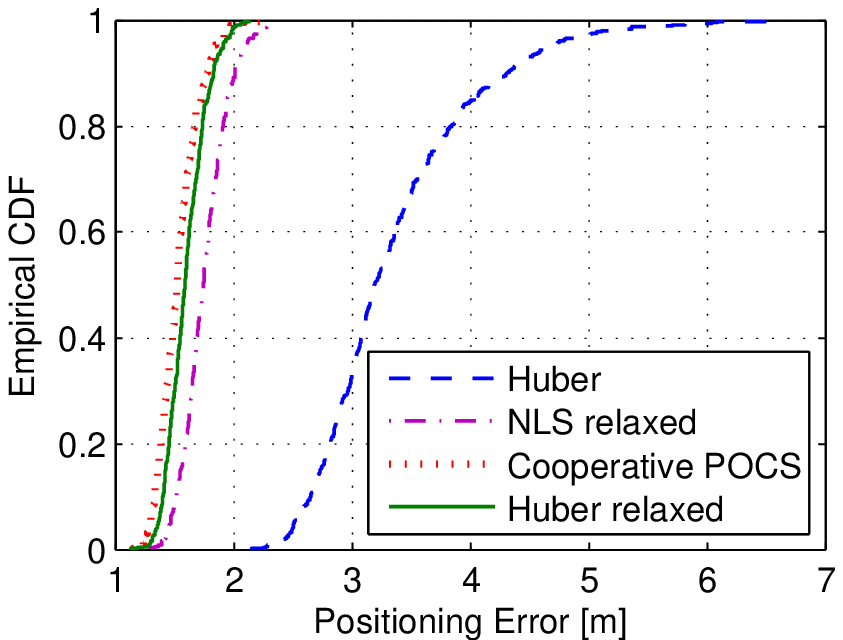}
}~
\subfloat[]{
\includegraphics[width=51mm,height=39mm]{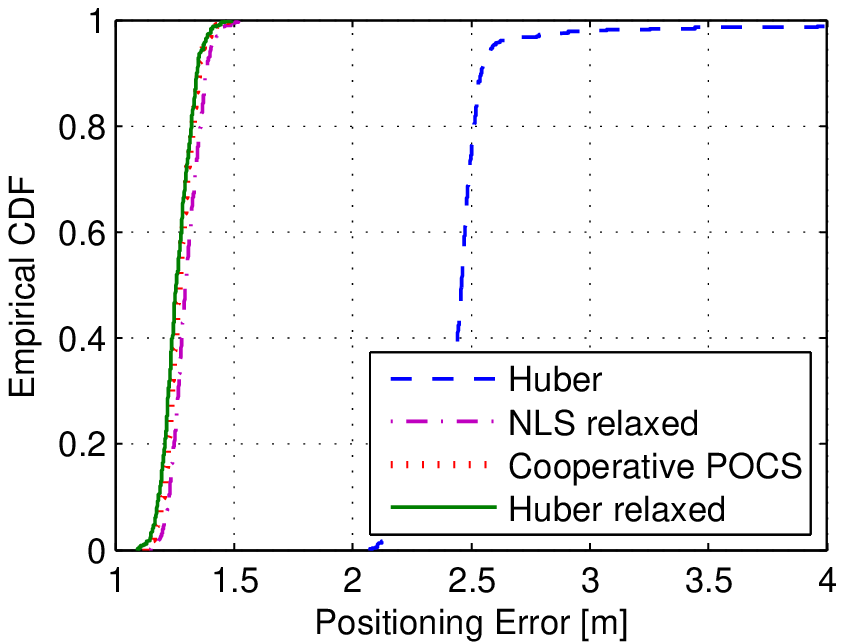}
}\\
\caption{CDF of different methods and the proposed Huber relaxation: (a) $P_{\cal N} =0.95$; (b) $P_{\cal N}=0.5$; (c) $P_{\cal N} =0.05$.}
\label{fig:CDF_Relaxation}
\end{figure*}

\begin{figure*}[htbp]
\centering
\subfloat[]{
\includegraphics[width=51mm,height=39mm]{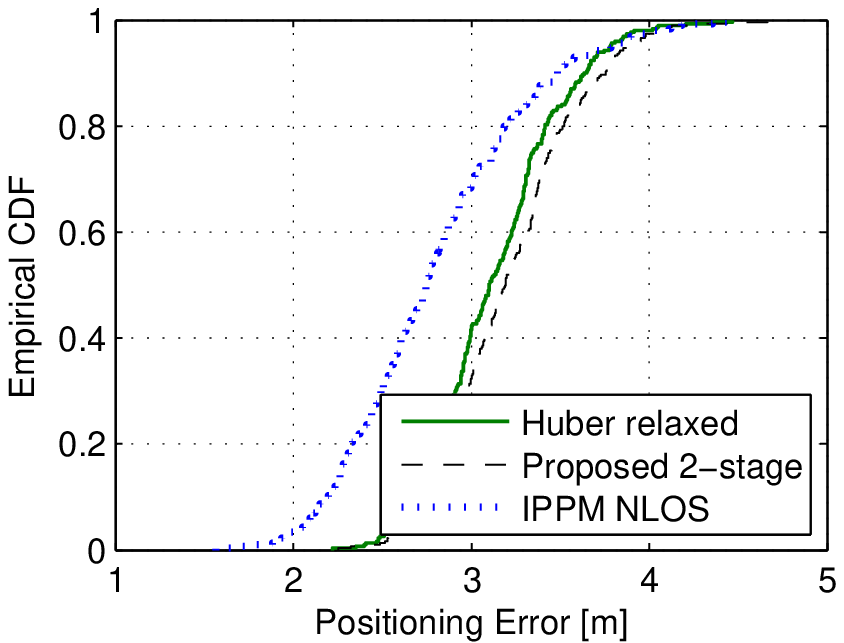}
}~
\subfloat[]{
\includegraphics[width=51mm,height=39mm]{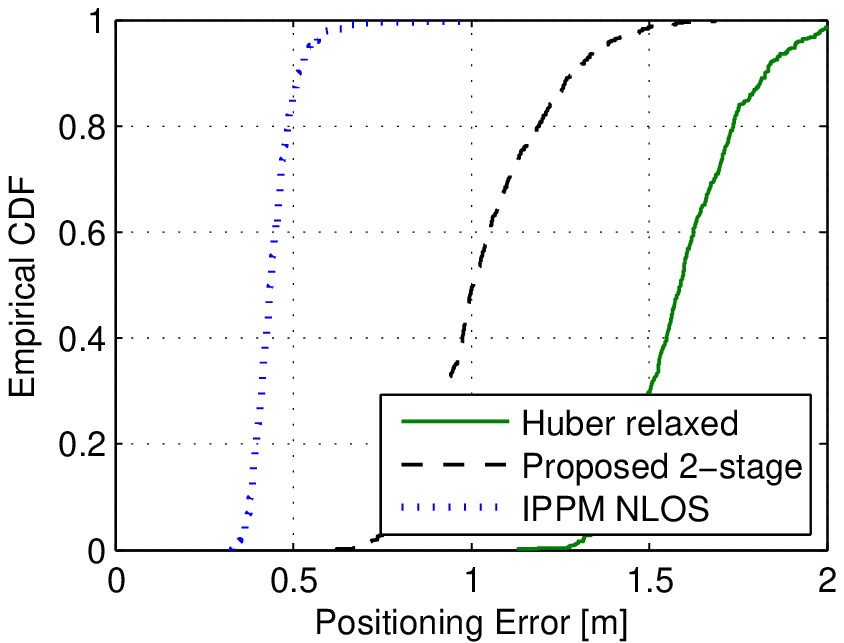}
}~
\subfloat[]{
\includegraphics[width=51mm,height=39mm]{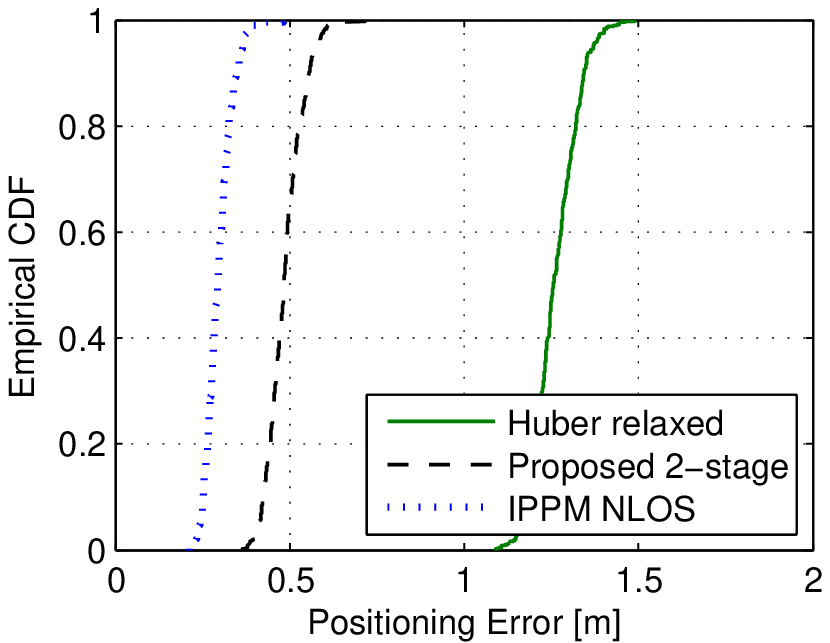}
}\\
\caption{CDF of the proposed 2-stage algorithm and the IPPM in \cite{Jia_IPPM_NLOS}: (a) $P_{\cal N} =0.95$; (b) $P_{\cal N}=0.5$; (c) $P_{\cal N} =0.05$.}
\label{fig:CDF_Full}
\end{figure*}

Selecting a suitable $K_2$ is very important at this stage as it enables a trade off between robustness and accuracy.
If the ratio of NLOS link is high, then selecting $K_2$ as done usually for Huber M-estimation, i.e., $1.5 \sigma_n \leq K_2 \leq 2 \sigma_n$, might even result in deterioration of the position estimates.
Thus, in this scenario, it is preferred to keep $\alpha_2$ very small, so the second stage does not change the position estimates obtained in the first stage.
On the other hand, if the ratio of the NLOS to LOS measurements is low, then the second stage can improve the positioning performance noticeably by selecting $1.5 \leq \alpha_2 \leq 2$.
We note that the estimation performance is still improved when a smaller value of $\alpha_2$ is chosen.
Therefore, if we have an \textit{a priori} estimate of the ratio of the NLOS to LOS measurements or the probability of a link being NLOS, then we can select $K_2$ according to the discussion above.
However, if such information is not available, then we should select small $\alpha_2$, e.g., $\alpha_2 = 0.1$, to achieve robust estimation result in every scenario. 

Although the second stage might improve the localization accuracy, it requires a number of iterations to converge, which increases the computational cost and communication load over the network.
Therefore, by tuning the stopping criteria in both stages of our algorithm we can have a trade off between computational cost and localization accuracy.

\section{Test and Validation} \label{Sec:Test}
\begin{figure*}[htbp] 
\centering
\subfloat[]{
\includegraphics[width=51mm,height=39mm]{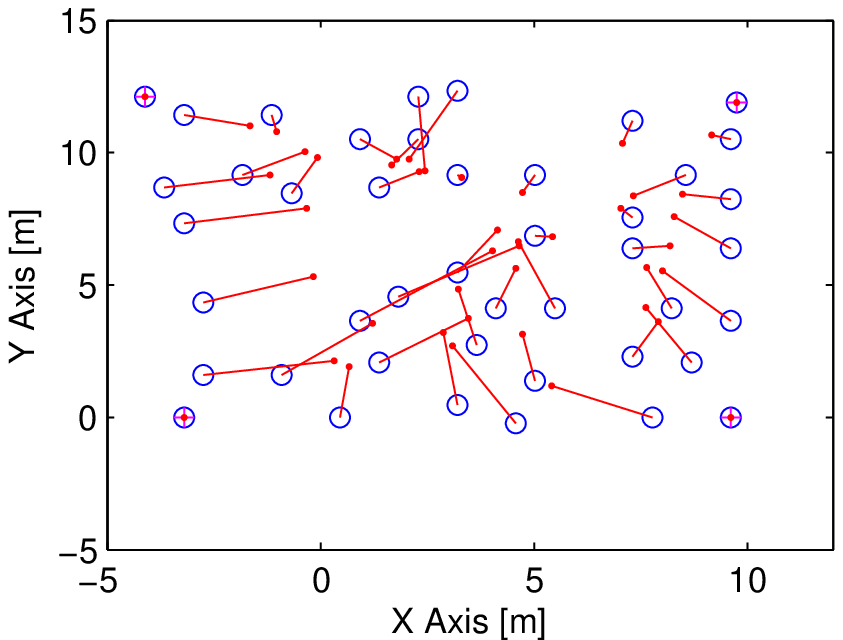}
}~
\subfloat[]{
\includegraphics[width=51mm,height=39mm]{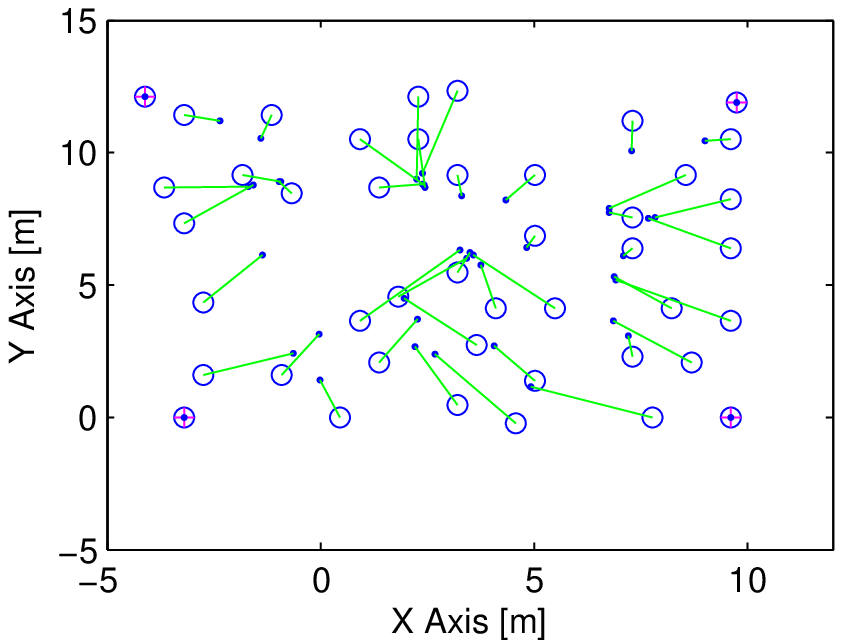}
\label{fig.:node_msd_trans_6nodes}
}~
\subfloat[]{
\includegraphics[width=51mm,height=39mm]{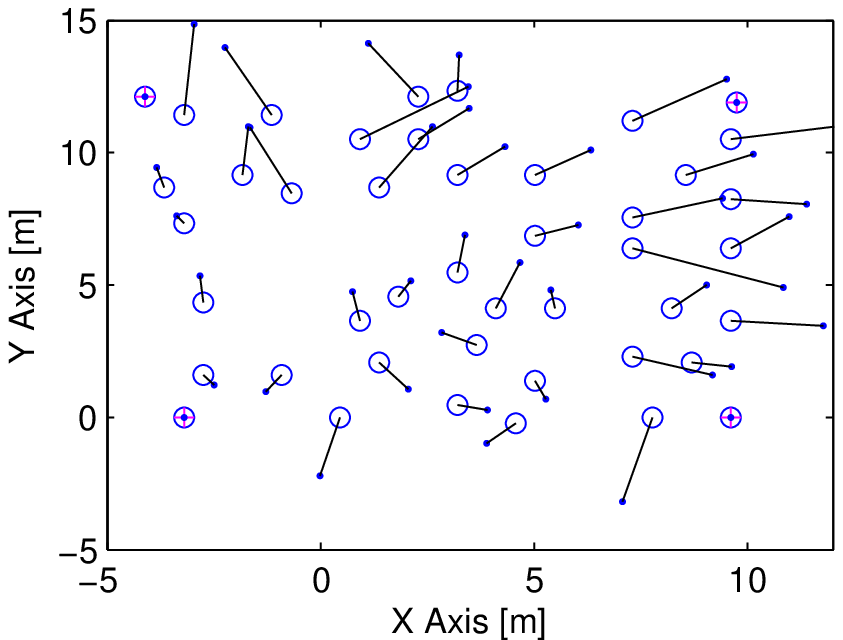}
}\\
\subfloat[]{
\includegraphics[width=51mm,height=39mm]{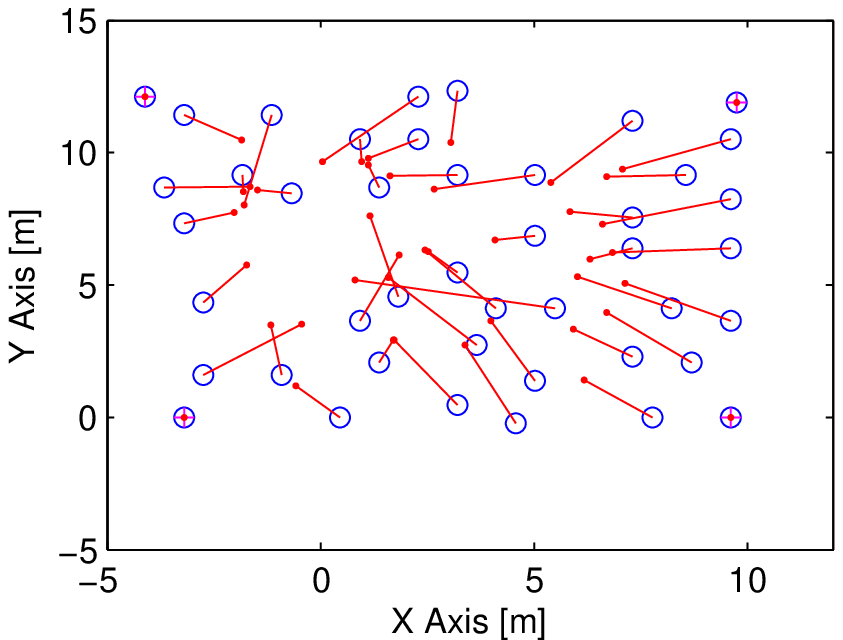}
}~
\subfloat[]{
\includegraphics[width=51mm,height=39mm]{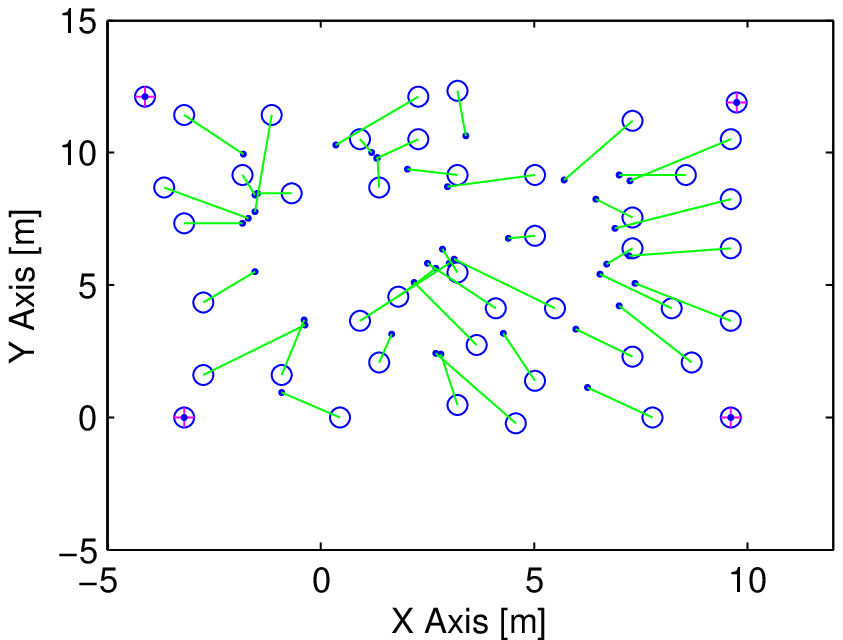}
}~
\subfloat[]{
\includegraphics[width=51mm,height=39mm]{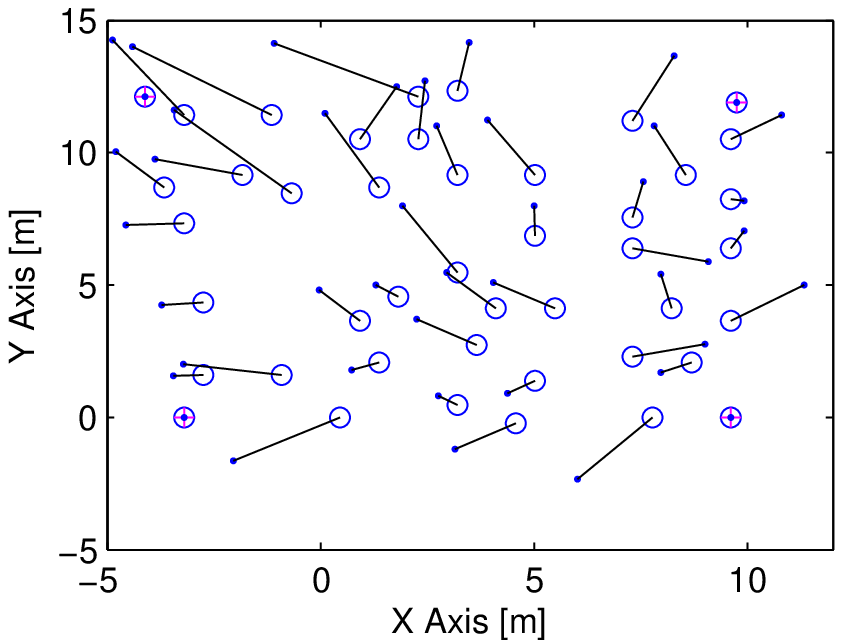}
}\\
\subfloat[]{
\includegraphics[width=51mm,height=39mm]{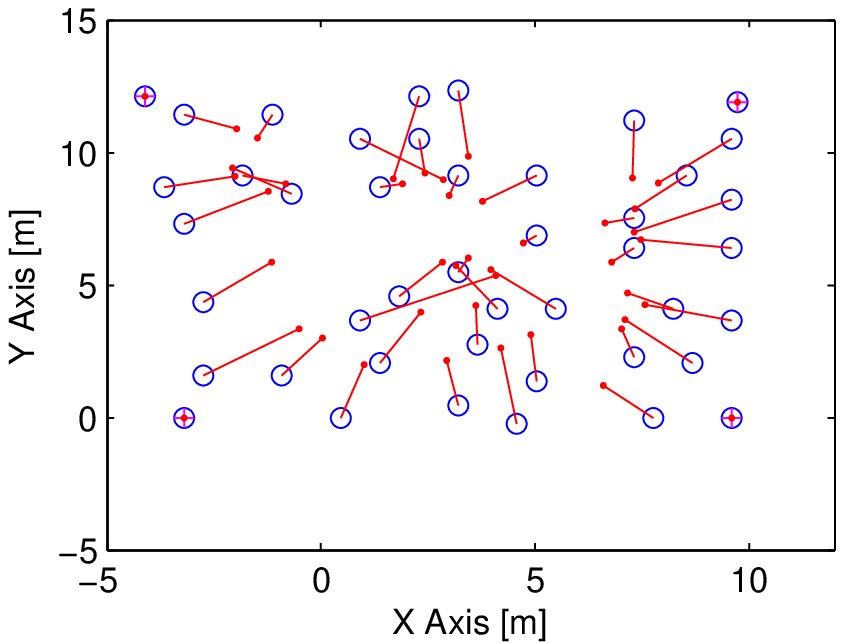}
}~
\subfloat[]{
\includegraphics[width=51mm,height=39mm]{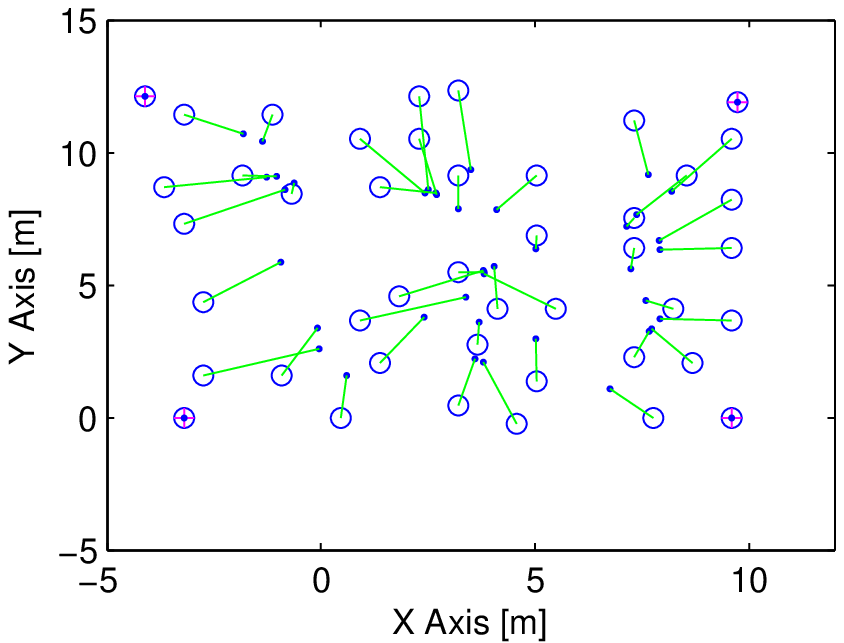}
}~
\subfloat[]{
\includegraphics[width=51mm,height=39mm]{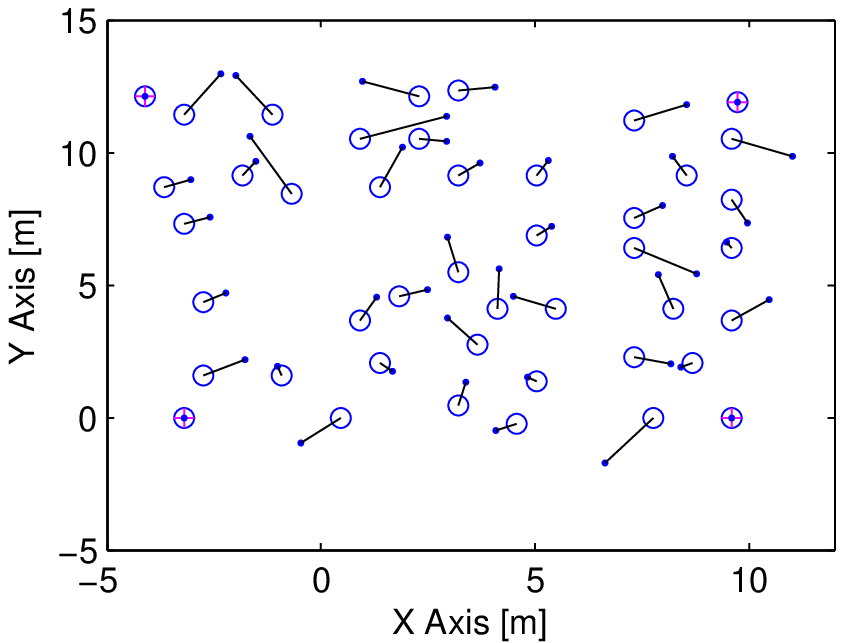}
}\\
\caption{Experimental data results. Localization performance with large $P_{\cal N}$: (a) cooperative POCS \cite{Gholami_POCS}; (b) first stage; (c) second stage.
Localization performance with moderate $P_{\cal N} $: (d) cooperative POCS \cite{Gholami_POCS}; (e) first stage; (f) second stage.
Localization performance with small $P_{\cal N}$: (g) cooperative POCS \cite{Gholami_POCS}; (h) first stage; (i) second stage.}
\label{fig.:Experimental_Nodes}
\end{figure*}

\subsection{Simulation Results} 
In this part, the performance of the proposed method is evaluated through simulations.
We consider a network of $M=4$ anchors and $N=50$ sensors located on a 2D space.
The sensors are randomly distributed on the plane while the anchors are at fixed locations $\x_{N+1}=[0 , 0]^T$, $\x_{N+2}=[10, 0 ]^T$, $\x_{N+3}=[10 , 10]^T$, and $\x_{N+4}=[0 , 10]^T$, where the units are in meters.
The range measurements were generated according to the model in \eqref{Range} with $\sigma_n=0.5$m and the NLOS bias is modelled as an exponential random variable with parameter $\gamma=10$m.
The Monte Carlo (MC) simulations are done under 500 runs.
As a performance metric, the network-average localization error for each noise realization, i.e., $ \sqrt{ \sum_{i=1}^{N}  \| \hat{\x}_i - \x_i  \|^2/N  }$ is evaluated.



We first consider the proposed convex relaxation and run this algorithm with $\mu_1=0.04$ and $K_1=2\sigma_n$.
We compare the proposed technique with the relaxation of the NLS in \cite{Abramo} with similar parameters and the same number of iterations, denoted by NLS relaxed.
We also apply the mentioned iterative technique with the same parameters and iteration number on the original Huber cost function and denote it by Huber.
Furthermore, we consider the cooperative POCS with parameter $\lambda^{l}=0$, thus it becomes almost similar to the IPPM in \cite{Jia_IPPM}, except that the projection is only implemented when $\|\x_i^{(l)} - \x_j^{(l)}\| \geq r_{ij}$.
The initial sensor positions for all algorithms are selected to have a Gaussian distribution with mean equal to the true sensor positions and standard deviation of 10 meters.
We define $P_{\cal N}$ as the probability of a link being NLOS.
We now consider three scenarios where the probability that a link is in NLOS is chosen to be  $P_{\cal N}=0.95$, $P_{\cal N}=0.5$, and  $P_{\cal N}=0.05$.
In Fig. \ref{fig:CDF_Relaxation}, the CDF of positioning error for different algorithms under various NLOS contamination level is shown after 50 iterations.
As observed in Fig.  \ref{fig:CDF_Relaxation}, the relaxation of Huber cost function is slightly better than the relaxation of NLS, and it has almost the same performance as POCS.
The original Huber cost function does not achieve a good result due to the lack of convexity and poor initialization.


To do further position refinement, we also simulate the second phase of our algorithm with $\mu_2=0.01$ and $K_2=0.1 \sigma_n$.
For the initialization, we use the position estimates obtained at the first stage by our proposed algorithm using convex relaxation of Huber cost function.
To have a lower bound on the performance of our algorithm, we implement the IPPM proposed in \cite{Jia_IPPM_NLOS} with the knowledge of perfect NLOS identification and denote it by IPPM NLOS.
Since the IPPM algorithm may not necessarily converge to a good solution because of lack of convexity, we use the position estimates obtained by cooperative POCS as initial points.
The CDF of the error of our 2-stage algorithm is illustrated in Fig. \ref{fig:CDF_Full} along with the IPPM with prior NLOS identification, where in both, 50 iterations are considered.
The CDF of the error of the proposed convex relaxation shown in Fig. \ref{fig:CDF_Relaxation} is also plotted in Fig. \ref{fig:CDF_Full}.
The results show that when the ratio of the NLOS to LOS measurements is high, the second stage of the algorithm might not improve the localization performance necessarily.
However, when the ratio of the NLOS to LOS links decreases, the second stage can improve the estimates obtained in the first stage distinguishably.
The performance of the proposed 2-stage algorithm is close to IPPM NLOS, which is based on perfect NLOS identification.




\subsection{Experimental Results}
\begin{figure*}[htbp]
\centering
\subfloat[]{
\includegraphics[width=51mm,height=39mm]{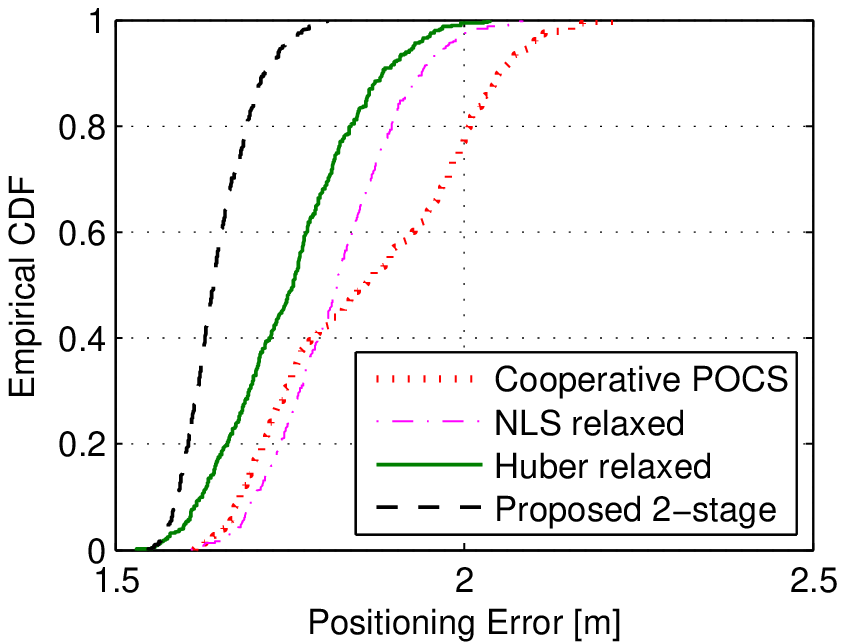}
}~
\subfloat[]{
\includegraphics[width=51mm,height=39mm]{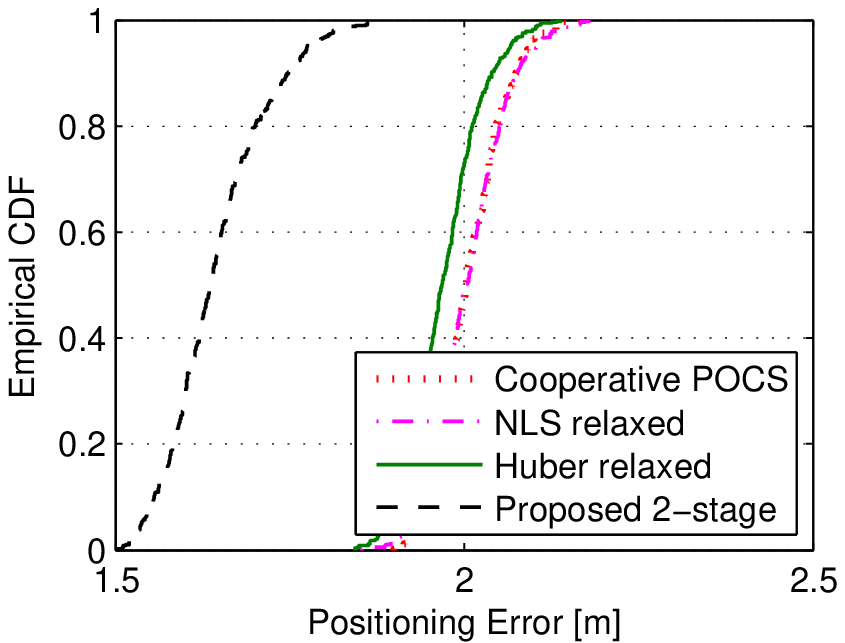}
}~
\subfloat[]{
\includegraphics[width=51mm,height=39mm]{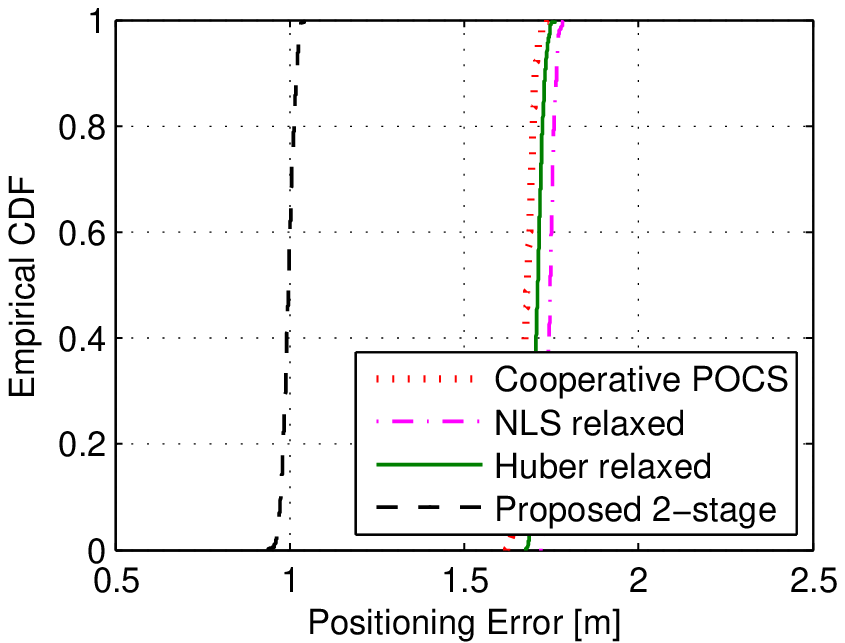}
}\\
\caption{The CDF of the positioning error using experimental data: (a) Large $P_{\cal N}$; (b) Moderate $P_{\cal N}$; (c) Small $P_{\cal N}$.}
\label{fig.:Experimental_MSE}
\end{figure*}
%
%
%
%
In this part, we consider localization of sensors using real data obtained by the measurement campaign reported in \cite{Patwari_Camp}.
The environment was an indoor office and there were 44 node locations where the transmitter and receiver were used at each location and pairwise range measurements were obtained.
We consider four nodes in the corner as anchor nodes with perfect location information and the other 40 nodes as the sensors with unknown locations.
Due to the scatterers and NLOS in the office, almost all of the measurements are affected by large positive errors as mentioned in \cite{Patwari_Camp}.
It is mentioned that the average amount of error is also calculated for these measurements, therefore, by subtracting that quantity from the measurements, a less unbiased set of measurement is obtained.
To evaluate the performance of our algorithm in different conditions we consider these scenarios:
\begin{itemize}
\item The raw measurements are considered, hence many of the measurements have positive errors, i.e., $P_{\cal N}$ is large.
\item The positive bias is subtracted from half of the measurements randomly, hence $P_{\cal N}$ is moderate.
\item The average bias is subtracted from all the raw measurements, thus $P_{\cal N} $ is small.
\end{itemize}
Using the unbiased measurements, the standard deviation of measurement noise is estimated roughly to be $\sigma_n=1$m.
By applying the iterative gradient descent technique on the proposed convex Huber cost function with $\mu_1=0.04$ and $K_1=2\sigma_n$, an estimate of the positions of sensors are obtained iteratively for 50 iterations.
The position estimates are also refined in the second stage with $K_2= 0.1 \sigma_n $ and $\mu_2=0.01$ for 50 iterations.
The final estimates at the end of each stage of our algorithm and the estimates obtained by cooperative POCS are shown along with the true sensor positions in Fig. \ref{fig.:Experimental_Nodes}.
The CDF of the positioning error in different NLOS scenarios are also illustrated in Fig. \ref{fig.:Experimental_MSE} by running 500 MC trials.

The results show that in general the relaxed Huber function achieves a better result compared to the other approaches.
Moreover, the second stage of the algorithm noticeably improves the position estimates obtained in the first stage, especially when $P_{\cal N}$ is small.


\section{Conclusion} \label{Sec:Conclusion} 
A robust distributed cooperative localization technique has been proposed in this work.
We first applied a convex relaxation on the Huber cost function and decent position estimates were obtained iteratively.
In the second stage of our algorithm, by iteratively minimizing the Huber loss function, it was shown that further refinement of position estimates could be generally obtained.
For iterative optimization in each stage, a gradient descent method was used.
By testing real data set, the superiority of our algorithm was verified.
We conclude that our 2-stage algorithm performs robustly against outliers; in particular it significantly outperforms other distributed techniques when the ratio of NLOS to LOS measurements is low.

\section*{Acknowledgment}
This work was supported by grants from the Natural Science and Engineering Research Council (NSERC) of Canada.



%




\bibliographystyle{IEEEtran}
\bibliography{References}

\end{document}